# Stretching and Shearing Contamination Analysis for Liutex/Rortex and Other Vortex Identification Methods

Pushpa Shrestha, Charles Nottage, Yifei Yu, Oscar Alvarez and Chaoqun Liu[1]

*Department of Mathematics, The University of Texas at Arlington, Arlington, Texas 76013, USA*

cliu@uta.edu

**Abstract:** Although traditional vortex-identification methods such as $Q$, $\Delta$, $\lambda_2$, and $\lambda_{ci}$ remain popular in the identification and visualization of vortices, these methods count on shearing and stretching as a part of vortex strength. However, shearing and stretching do not contribute to fluid rotation. In this paper, the contamination effects of stretching and shearing of these methods are investigated and compared with Liutex method. From our investigation, the Liutex is an exact definition of fluid rotation or vortex, while other vortex identification methods are contaminated by stretching and shearing at different levels. The decomposition of the velocity gradient tensor can only be conducted in a so-called Principal Coordinate for uniqueness. The mathematical relation between Liutex/Rortex and other vortex identification function are derived in this paper and then the effects of shearing and stretching on different vortex identification methods are studied. The mathematical formula and computation of the stretching and shearing effects on different schemes clearly show that the Liutex method has superiority over the other vortex identification methods as it only counts on local fluid rigid rotation while other methods count on stretching/compression and shearing as a part of the fluid rotation or vortex.

**Keywords**: Liutex, vorticity, vortex, Omega method, transition flow, velocity gradient tensor, principal coordinate, principal tensor, Liutex tensor decomposition.

## 1. Introduction

Vortices can be observed everywhere in nature; one of the examples is turbulence. Turbulence is formed with countless vortices of different sizes and strengths in the fluid flow field. A vortex is naturally recognized as the rotational motion of fluids. Within the last several decades, many vortex identification methods have been developed to define, analyze, and visualize the vortex structure in the flow field; however, we still lack unambiguous and universally accepted vortex definition method. This lack of clarity is the impediment that has caused many confusions and misunderstandings in turbulence research (Liu et al., 2014)[1]. According to Lugt (1979)[2] and Robinson (1990)[3], an intuitive existence of a vortex can be verified by closed or spiraling streamlines or path lines in fluid flow. However, the streamlines or path lines are not Galilean invariant and thus cannot be a proper way for vortex identification. A critical theory was proposed by Chong and Perry (1990)[4] to judge the presence of the local rotational motion. If the velocity gradient tensor of a three-dimensional flow field has a pair of complex conjugate eigenvalues and a real eigenvalue, the instantaneous streamline pattern presents a local swirling motion around the axis of the local fluid rotation, which is the real eigenvector of velocity gradient tensor and also known as direction of Liutex (Liu et al.) [7]. The concept of the vorticity concentration and other vorticity-based methods (Helmholtz, H.; Saffma, P., 1990)[5-6], was widely accepted by many researchers since the vorticity vector was thought to provide a mathematical definition of the fluid rotational motion. These vorticity-based methods are classified as the first generation (1G) of vortex identification methods, according to Liu et al. (2019)[7]. Zhou and Antonia (1993)[8] utilized the spatially phase correlated vorticity to characterize large-scale and organized structures in the cylinder wake; however, some serious problems arise in most viscous flows, particularly in turbulent flows. The reality is that vorticity cannot distinguish a vortical region with rotational motions from a strong shear layer (Epps, B., 2017)[9]. With this issue in mind, several scalar-based vortex identification criteria, such as $Q$ (Hunt et al., 1988)[10], $\lambda_2$ (Jeong, J., and Hussain, F., 1995)[11], $\lambda_{ci}$ (Zhou et al., 1999)[12], $\Omega$ (Liu et al., 2016; Dong et al., 2018)[13-14], etc., have been prompted and extensively applied in visualization of vortex structure. These criteria are regarded as the second generation (2G) of vortex identification methods by Liu et al. (2019)[7]. However, $Q$ and $\lambda_2$ are restrictive for vortex identification for incompressible flows due to their incompressibility assumption.

---

[1] **Corresponding author:** Chaoqun Liu, Email: cliu@uta.edu

Additionally, for most scalar-based vortex identification criteria, the sensitivity to the threshold selection provides a large amount of difficulties in making a judgment to define the boundary of vortical structures (Liu et al., 2016; Zhang et al., 2018)[13, 15]. These issues prompted the development of the new concept of Liutex/Rortex (Liu et al., 2018; Gao et al., 2018)[16-17]. Unlike its predecessors, the Liutex method is a novel eigenvector-based method which is local, accurate, unique, and systematically defined. Furthermore, the systematical definition of Liutex is given in scalar, vector, and tensor forms[17]. Liutex is defined as a vector. The scalar form of Liutex represents the rotational strength of the fluid rotation, whereas the vector form gives the direction of the local fluid rotation. The Liutex vector is also Galilean invariant (Haller, 2005; Wang et al., 2018; Liu et al., 2019)[18-20]. The Liutex method, Liutex-Omega method (Liu et al., 2018)[16], Liutex-Core – Line method and other Liutex-based methods are regarded as the third generation (3G) of vortex identification methods, according to Liu et al. [7] and countless Liutex users. In this paper, we demonstrate that the shear and stretching contamination of several 2G vortex identification methods from both mathematical formula and numerical experiments. A velocity vector at some points in our numerical domain (flow field) is chosen, and then a velocity gradient tensor is calculated at that point. After that, we decompose the velocity gradient tensor at that point in the flow field based on the Liutex Principal Decomposition, which is unique and is done in the Principal Coordinate. In other words, the velocity gradient tensor is uniquely decomposed into rigid rotation part, stretching part, and shearing part in the Principal Coordinate, which is explained in section 3 in detail.

This paper is organized as follows. In Section 2, some traditional vortex identification methods are revisited including $\Delta$, $Q$, $\lambda_2$ and $\lambda_{ci}$ methods. Section 3 describes the Liutex-based Principal Decomposition in the unique Principal Coordinate. Based on the unique Liutex tensor decomposition in the Principal Coordinate, the theoretical contamination analysis of $\Delta$, $Q$, $\lambda_2$, $\lambda_{ci}$ and $Liutex$ schemes is given in Section 4. Numerical experiments' and results can be founded in Sections 5 and 6 respectively, and some concluding remarks are made in the last section.

## 2. Review of Some Traditional Vortex Identification Methods.

### 2.1 $\Delta$ Method

The $\Delta$ method defines a vortex to be the region where the velocity gradient tensor $\vec{\nabla v}$ has a pair of complex conjugate eigenvalues and a real eigenvalue (Chong et al., 1990)[21]. If $\lambda_1, \lambda_2$ and $\lambda_3$ are three eigenvalues of the 3× 3 matrix of $\vec{\nabla v}$, then the characteristic equation can be written as:

$$\lambda^3 + I_1\lambda^2 + I_2\lambda + I_3 = 0 \tag{1}$$

where, $I_1$, $I_2$, and $I_3$ are the first, second and third invariants of the characteristic equation (1) and given by

$$I_1 = -(\lambda_1 + \lambda_2 + \lambda_3) = -tr(\boldsymbol{\nabla \vec{v}}) \tag{2}$$

$$I_2 = \lambda_1\lambda_2 + \lambda_2\lambda_3 + \lambda_3\lambda_1 = -\frac{1}{2}[tr(\boldsymbol{\nabla \vec{v}}^2) - tr(\boldsymbol{\nabla \vec{v}})^2] \tag{3}$$

$$I_3 = -\lambda_1\lambda_2\lambda_3 = -\det(\boldsymbol{\nabla \vec{v}}) \tag{4}$$

The discriminant of the characteristic equation (1) of the velocity gradient tensor is given by

$$\Delta = \left(\frac{\tilde{Q}}{3}\right)^3 + \left(\frac{\tilde{R}}{2}\right)^2 \tag{5}$$

where, $\tilde{Q} = I_2 - \frac{1}{3}I_1^2$ and, $\tilde{R} = I_3 + \frac{2}{27}I_1^3 - \frac{1}{3}I_1I_2$.

For incompressible flow, the first invariant $I_1 = 0$. So, in this case, the discriminant of the characteristic equation (5) becomes $\Delta = \left(\frac{I_2}{3}\right)^3 + \left(\frac{I_3}{2}\right)^2$. If $\Delta \leq 0$, all three eigenvalues of $\boldsymbol{\nabla \vec{v}}$ are real, but if $\Delta > 0$, there exists one real and two conjugate complex eigenvalues. The later means that the point is inside a vortex region. $\Delta$ method is a scalar method by using the iso-surface to show the vortex structure. The iso-surface is very sensitive to the selection of threshold, which is man-made and arbitrary in general.

## 2.2 Q Method

The $Q$ method is one of the most popular methods to visualize the vortex structure, which was proposed by Hunt et al. (1988)[10]. Although $Q$ is the second invariant in the eigenvalue equation (1) given above, the value of $Q$ can be calculated as half of the difference of squares of the Frobenius norm of vorticity tensor and strain-rate tensor. i.e.,

$$Q = \frac{1}{2}(\|\boldsymbol{B}\|_F^2 - \|\boldsymbol{A}\|_F^2) \tag{6}$$

where **A** and **B** are the symmetric and anti-symmetric part of the velocity gradient tensor, respectively.

$$A = \frac{1}{2}(\nabla \vec{v} + \nabla \vec{v}^T) = \begin{bmatrix} \frac{\partial u}{\partial x} & \frac{1}{2}\left(\frac{\partial u}{\partial y} + \frac{\partial v}{\partial x}\right) & \frac{1}{2}\left(\frac{\partial u}{\partial z} + \frac{\partial w}{\partial x}\right) \\ \frac{1}{2}\left(\frac{\partial v}{\partial x} + \frac{\partial u}{\partial y}\right) & \frac{\partial v}{\partial y} & \frac{1}{2}\left(\frac{\partial v}{\partial z} + \frac{\partial w}{\partial y}\right) \\ \frac{1}{2}\left(\frac{\partial w}{\partial x} + \frac{\partial u}{\partial z}\right) & \frac{1}{2}\left(\frac{\partial w}{\partial y} + \frac{\partial v}{\partial z}\right) & \frac{\partial w}{\partial z} \end{bmatrix} \quad (7)$$

$$B = \frac{1}{2}(\nabla \vec{v} - \nabla \vec{v}^T) = \begin{bmatrix} 0 & \frac{1}{2}\left(\frac{\partial u}{\partial y} - \frac{\partial v}{\partial x}\right) & \frac{1}{2}\left(\frac{\partial u}{\partial z} - \frac{\partial w}{\partial x}\right) \\ \frac{1}{2}\left(\frac{\partial v}{\partial x} - \frac{\partial u}{\partial y}\right) & 0 & \frac{1}{2}\left(\frac{\partial v}{\partial z} - \frac{\partial w}{\partial y}\right) \\ \frac{1}{2}\left(\frac{\partial w}{\partial x} - \frac{\partial u}{\partial z}\right) & \frac{1}{2}\left(\frac{\partial w}{\partial y} - \frac{\partial v}{\partial z}\right) & 0 \end{bmatrix} \quad (8)$$

The $Q$ method considers that a vortex occurs in the region where the second invariant of the characteristic equation (1) is positive, i.e., $Q > 0$. Using $Q$ method, it is easy to track the vortical structure by choosing iso-surface. However, $Q$ is a scalar, and a proper threshold is required to visualize the vortex region. Also, there does exist an inconsistency between the $\Delta$ method and $Q$ method as $\left(\frac{\tilde{R}}{2}\right)^2$ is always positive and even if $Q < 0$, it is still possible that $\Delta$ could be positive, which indicates the point is still inside a vortex region.

### 2.3 $\lambda_{ci}$ Criterion

The $\lambda_{ci}$ criterion (Zhou et al., 1999; Chakraborty et al., 2005)[12, 22] uses the imaginary part of the complex eigenvalues of the velocity gradient tensor to visualize the vortex structure. It is based on the idea that the local time-frozen streamlines exhibit a rotational flow pattern when $\nabla \vec{v}$ has a pair of complex conjugate eigenvalues. In this case, the tensor transformation of $\nabla \vec{v}$ is given by

$$\nabla \vec{v} = [\vec{v}_r \; \vec{v}_{cr} \; \vec{v}_{ci}] \begin{bmatrix} \lambda_r & 0 & 0 \\ 0 & \lambda_{cr} & \lambda_{ci} \\ 0 & -\lambda_{ci} & \lambda_{cr} \end{bmatrix} [\vec{v}_r \; \vec{v}_{cr} \; \vec{v}_{ci}]^{-1} \quad (9)$$

where $\lambda_r$ is the real eigenvalue with corresponding eigenvector $\vec{v}_r$ and the complex conjugate pair of complex eigenvalues are $\lambda_{cr} \pm i\lambda_{ci}$ with corresponding eigenvectors $\vec{v}_{cr} \pm i\,\vec{v}_{ci}$. In this case, in the local curvilinear system $(c_1, c_2, c_3)$ spanned by the eigenvector $(\vec{v}_r, \vec{v}_{cr}, \vec{v}_{ci})$, the instantaneous streamlines exhibit spiral motion. The equations of such streamlines can be written as:

$$c_1(t) = c_1(0)e^{\lambda_r t} \quad (10)$$

$$c_2(t) = [c_2(0)\cos(\lambda_{ci}t) + c_3(0)\sin(\lambda_{ci}t)]e^{\lambda_{cr}t} \quad (11)$$

$$c_3(t) = [c_3(0)\cos(\lambda_{ci}t) - c_2(0)sin(\lambda_{ci}t)]e^{\lambda_{cr}t} \qquad (12)$$

The strength of this swirling motion was improperly quantified by $\lambda_{ci}$ as the $\lambda_{ci}$ cannot be exactly the pure rotation strength. $\lambda_{ci}$ is also characterized as a scalar-valued criterion (Liu et al., 2019)[7]. Note that Eq. (9) represents a similar transformation (not orthogonal transformation) that cannot keep vorticity constant.

## 2.4 $\lambda_2$ criterion

The $\lambda_2$ criterion is calculated based on the observation that, in a vortex, pressure tends to be the local minimum on the axis of a swirling motion of fluid particles when the centrifugal force is balanced by the pressure force (the so-called cyclostrophic balance). It is valid only in a steady inviscid planar flow (Jeong, J., and Hussain, F., 1995)[23]. However, this assumption fails to identify vortices under strong unsteady and viscous effects accurately. By neglecting these unsteady and viscous effects, the symmetric part **S** of the gradient of the incompressible Navier–Stokes equation can be expressed as:

$$\mathbf{S} = \mathbf{A}^2 + \mathbf{B}^2 = -\frac{\nabla(\nabla p)}{\rho} \qquad (13)$$

where $p$ is the pressure and equation (13) is a representation of the pressure Hessian matrix $((\nabla(\nabla p))_{ij} = \frac{\partial^2 p}{\partial x_i \partial y_i}$. To capture the region of local pressure minimum in a plane perpendicular to vortex core line, Jeong & Hussain defined the vortex core as a connected region with two positive eigenvalues of the pressure Hessian matrix, i.e., a connected region with two negative eigenvalues of the symmetric tensor **S**. If $\lambda_{S1}, \lambda_{S2}$ & $\lambda_{S3}$ are three real eigenvalues of the symmetric tensor S and when setting them in order in such a way that $\lambda_{S1} \geq \lambda_{S2} \geq \lambda_{S3}$, there must be $\lambda_{S2} < 0$ as two eigenvalues are negative, which confirms that there exists vortex. In general, $\lambda_{S2}$ cannot be expressed in terms of eigenvalues of velocity gradient tensor; however, in some special cases when eigenvectors are orthonormal, $\lambda_{S2}$ can be exclusively determined by eigenvalues of velocity gradient tensor. Vortex structure can be visualized as iso-surface by selecting a proper threshold of $\lambda_{S2}$ (Liu et al., 2019)[7]. The relation between the eigenvalues of the symmetric tensor $\mathbf{A}^2+\mathbf{B}^2$ and second invariant $Q$ is given by:

$$Q = -\frac{1}{2}tr(\boldsymbol{A}^2 + \boldsymbol{B}^2) = -\frac{1}{2}(\lambda_{S1} + \lambda_{S2} + \lambda_{S3}) \tag{14}$$

It can be shown that while the $Q$ criterion measures the excess of vorticity rate over the strain rate magnitude in all directions, the $\lambda_2$ criterion looks for this excess only on a specific plane (Jeong, J., and Hussain, F., 1995)[23]. Another similar article, "on the relationship between vortex identifications methods" by Chakraborty & Balachandar (2005)[22] is also recommended. There is another article by Liu et al., "Third generation of vortex identification methods", which reviews the second generation of vortex identification methods – methods based on eigenvalues of $\nabla\vec{v}$ before presenting novel third generation of vortex identification methods like Omega and Liutex schemes[7]. Again, all of the above second generation methods for vortex identification are scalar and the vortex structure shown by these criteria is strongly dependent on the selection of threshold which is arbitrary.

**2.5 Liutex**

**Definition 1**: Liutex is a vector defined as

$$\boldsymbol{R} = R\boldsymbol{r}$$

Where R is the magnitude of Liutex, and $\boldsymbol{r}$ is the direction of Liutex.

According to Ref. [24] and [28], $\boldsymbol{r}$ is the real eigenvector of the velocity gradient tensor, and the explicit formula of R is

$$R = \vec{\omega}\cdot\vec{r} - \sqrt{(\vec{\omega}\cdot\vec{r})^2 - 4\lambda_{ci}^2} \tag{15}$$

Dr. Liu classified Liutex method as the third generation of vortex identification methods, while the Q method, $\lambda_{ci}$ criterion and $\lambda_2$ criterion are all classified as the second generation methods, since they are all scalar and eigenvalue related. Liutex is a vector which overcomes all the drawbacks of a scalar, e.g. a threshold is needed when drawing graphs. Liutex is a mathematical definition of fluid rotation or vortex.

## 3. Principal Coordinates and Principal Decomposition.

***Definition 2***: Principal Coordinate at a point is a coordinate that satisfies (Yu et al, 2020)[32]:

1. Its z-axis is parallel to the $\boldsymbol{r}$(direction of Liutex)

2. The velocity gradient tensor under this coordinate is in the form of:

$$\nabla V = \begin{bmatrix} \lambda_{cr} & \frac{\partial U}{\partial Y} & 0 \\ \frac{\partial V}{\partial X} & \lambda_{cr} & 0 \\ \frac{\partial W}{\partial X} & \frac{\partial W}{\partial Y} & \lambda_r \end{bmatrix} \quad (16)$$

Where $\lambda_r$ and $\lambda_{cr}$ are real eigenvalue and real part of the conjugate complex eigenvalue pair of the velocity gradient tensor respectively for rotation points.

3. $\frac{\partial U}{\partial Y} < 0$ and $\left|\frac{\partial U}{\partial Y}\right| < \left|\frac{\partial V}{\partial X}\right|$

***Theorem* 1**: Under Principal Coordinate, $\frac{\partial U}{\partial Y} = -\frac{R}{2}$, where R is the magnitude of Liutex and $\frac{\partial U}{\partial Y}$ is the component in (16)

***Proof***: We use the definition of Liutex magnitude in Ref.[7] to prove this theorem.

Given an arbitrary velocity gradient tensor $\nabla v$, there always exists a rotation matrix $Q_r$ which aligns Z axis of new frame XYZ with Liutex direction $\boldsymbol{r}$ after the rotation.

$$\nabla V = Q_r \nabla v Q_r^T = \begin{bmatrix} \frac{\partial U}{\partial X} & \frac{\partial U}{\partial Y} & 0 \\ \frac{\partial V}{\partial X} & \frac{\partial V}{\partial Y} & 0 \\ \frac{\partial W}{\partial X} & \frac{\partial W}{\partial Y} & \frac{\partial W}{\partial Z} \end{bmatrix} \quad (17)$$

Then, a second rotation $P_r$ around Z-axes (P rotation) is applied.

$$P_r = \begin{bmatrix} \cos\theta & \sin\theta & 0 \\ -\sin\theta & \cos\theta & 0 \\ 0 & 0 & 1 \end{bmatrix} \quad (18)$$

And, the new velocity gradient tensor $\nabla V_\theta$ after rotation is:

$$\nabla V_\theta = P_r \nabla V P_r^T = \begin{bmatrix} \frac{\partial U}{\partial X}\big|_\theta & \frac{\partial U}{\partial Y}\big|_\theta & 0 \\ \frac{\partial V}{\partial X}\big|_\theta & \frac{\partial V}{\partial Y}\big|_\theta & 0 \\ \frac{\partial W}{\partial X}\big|_\theta & \frac{\partial W}{\partial Y}\big|_\theta & \frac{\partial W}{\partial Z}\big|_\theta \end{bmatrix} \qquad (19)$$

Where(Liu et al, 2019)[7]

$$\frac{\partial U}{\partial Y}\bigg|_\theta = \alpha \sin(2\theta + \varphi) - \beta \qquad (20)$$

$$\frac{\partial V}{\partial X}\bigg|_\theta = \alpha \sin(2\theta + \varphi) + \beta \qquad (21)$$

$$\frac{\partial U}{\partial X}\bigg|_\theta = -\alpha \cos(2\theta + \varphi) + \frac{1}{2}\left(\frac{\partial U}{\partial X} + \frac{\partial V}{\partial Y}\right) \qquad (22)$$

$$\frac{\partial V}{\partial Y}\bigg|_\theta = \alpha \cos(2\theta + \varphi) + \frac{1}{2}\left(\frac{\partial U}{\partial X} + \frac{\partial V}{\partial Y}\right) \qquad (23)$$

$$\alpha = \frac{1}{2}\sqrt{\left(\frac{\partial v}{\partial y} - \frac{\partial u}{\partial x}\right)^2 + \left(\frac{\partial v}{\partial x} + \frac{\partial u}{\partial y}\right)^2} \qquad (24)$$

$$\beta = \frac{1}{2}\left(\frac{\partial v}{\partial x} - \frac{\partial u}{\partial y}\right) \qquad (25)$$

$$\varphi = \begin{cases} \arctan\left(\frac{\frac{\partial v}{\partial x} + \frac{\partial u}{\partial y}}{\frac{\partial v}{\partial y} - \frac{\partial u}{\partial x}}\right), & \frac{\partial v}{\partial y} - \frac{\partial u}{\partial x} \neq 0 \\ \frac{\pi}{2}, & \frac{\partial v}{\partial y} - \frac{\partial u}{\partial x} = 0, \frac{\partial v}{\partial x} + \frac{\partial u}{\partial y} > 0 \\ -\frac{\pi}{2}, & \frac{\partial v}{\partial y} - \frac{\partial u}{\partial x} = 0, \frac{\partial v}{\partial x} + \frac{\partial u}{\partial y} < 0 \end{cases} \qquad (26)$$

Then, the Liutex magnitude is defined as (Liu et al, 2019)[7]

$$R = \begin{cases} 2(|\beta| - \alpha), & \beta^2 > \alpha^2 \\ 0, & o.w. \end{cases} \qquad (27)$$

Since we are interested in points which are inside the vortex boundary, so it is assumed $\beta^2 > \alpha^2$, otherwise there is no vortex.

From part (2) in Def. 2,

$$\left.\frac{\partial U}{\partial X}\right|_\theta = \left.\frac{\partial V}{\partial Y}\right|_\theta \tag{28}$$

Thus,

$$\cos(2\theta + \varphi) = 0 \tag{29}$$

Then, $\sin(2\theta + \varphi) = 1 \text{ or } -1$.

Case 1: $\beta > 0$ and $\sin(2\theta + \varphi) = 1$

$$\left.\frac{\partial U}{\partial Y}\right|_\theta = \alpha - \beta = -\frac{1}{2}R \tag{30}$$

Case 2: $\beta > 0$ and $\sin(2\theta + \varphi) = -1$

$$\left.\frac{\partial U}{\partial Y}\right|_\theta = -\alpha - \beta \tag{31}$$

$$\left.\frac{\partial V}{\partial X}\right|_\theta = -\alpha + \beta \tag{32}$$

However,

$$|-\alpha - \beta| > |-\alpha + \beta| \tag{33}$$

It violates the part (3) in Def. 2, and as a result $\sin(2\theta + \varphi) \neq -1$

Case 3: $\beta < 0$ and $\sin(2\theta + \varphi) = 1$

$$\left.\frac{\partial U}{\partial Y}\right|_\theta = \alpha - \beta \tag{34}$$

$$\left.\frac{\partial V}{\partial X}\right|_\theta = \alpha + \beta \tag{35}$$

However,

$$|\alpha - \beta| > |\alpha + \beta| \tag{36}$$

The part (3) in Def.2 is not satisfied, so $\sin(2\theta + \varphi) \neq 1$

Case 4: $\beta < 0$ and $\sin(2\theta + \varphi) = -1$

$$\left.\frac{\partial U}{\partial Y}\right|_\theta = \alpha - \beta = -\frac{1}{2}R \tag{37}$$

Theorem 1 is proved.

**Theorem 2**: For an arbitrary velocity gradient tensor, there always exists one and only one Principal Coordinate.

Proof: A coordinate has x, y and z axes. If directions of these 3 axes are determined, the coordinate is well-defined. Firstly, the direction of z-axes is unique, defined as $r$ (direction of Liutex). All the coordinates satisfying the z-axis requirement can be achieved by rotating coordinate around $r$. Therefore, if the rotation angle is determined, so is the coordinate.

In the proof of theorem 1, if $\beta > 0$, then $\cos(2\theta + \varphi)$ and $\sin(2\theta + \varphi)$ must be 0 and 1, respectively. Thus, $2\theta + \varphi = \frac{\pi}{2} \Rightarrow \theta = \frac{1}{2}(\frac{\pi}{2} - \varphi)$. Similarly, if $\beta < 0$, then $\cos(2\theta + \varphi)$ and $\sin(2\theta + \varphi)$ must be 0 and -1. So, $2\theta + \varphi = \frac{3\pi}{2} \Rightarrow \theta = \frac{1}{2}(\frac{3\pi}{2} - \varphi)$.

In conclusion, the Principal Coordinate is unique.

**Definition 3**: The Principal Decomposition is the decomposition under the Principal Coordinate i.e.

$$\nabla V = \begin{bmatrix} \lambda_{cr} & -\frac{R}{2} & 0 \\ \frac{R}{2}+\varepsilon & \lambda_{cr} & 0 \\ \xi & \eta & \lambda_r \end{bmatrix} = \begin{bmatrix} 0 & -\frac{R}{2} & 0 \\ \frac{R}{2} & 0 & 0 \\ 0 & 0 & 0 \end{bmatrix} + \begin{bmatrix} 0 & 0 & 0 \\ \varepsilon & 0 & 0 \\ \xi & \eta & 0 \end{bmatrix} + \begin{bmatrix} \lambda_{cr} & 0 & 0 \\ 0 & \lambda_{cr} & 0 \\ 0 & 0 & \lambda_r \end{bmatrix} = A + B + C \tag{38}$$

Where A represents the rotation part, B represents the shear part, and C represents the stretching part.

The Principal Decomposition correctly and uniquely decomposes the velocity gradient tensor into the rotation part, shear part, and stretching part. So, the shear and stretching contamination analysis can be done uniquely only in the Principal Coordinate[32].

## 4. Theoretical Contamination Analysis

Suppose the velocity gradient tensor in the Principal Coordinate is:

$$\nabla \vec{V} = \begin{bmatrix} \lambda_{cr} & -\frac{1}{2}R & 0 \\ \frac{1}{2}R + \varepsilon & \lambda_{cr} & 0 \\ \xi & \eta & \lambda_r \end{bmatrix} \tag{39}$$

The Principal Decomposition, which is the decomposition of velocity gradient tensor in the Principal Coordinate, is given by

$$\nabla \vec{V} = \begin{bmatrix} \lambda_{cr} & -\frac{1}{2}R & 0 \\ \frac{1}{2}R + \varepsilon & \lambda_{cr} & 0 \\ \xi & \eta & \lambda_r \end{bmatrix} = \begin{bmatrix} 0 & -\frac{R}{2} & 0 \\ \frac{R}{2} & 0 & 0 \\ 0 & 0 & 0 \end{bmatrix} + \begin{bmatrix} 0 & 0 & 0 \\ \varepsilon & 0 & 0 \\ \xi & \eta & 0 \end{bmatrix} + \begin{bmatrix} \lambda_{cr} & 0 & 0 \\ 0 & \lambda_{cr} & 0 \\ 0 & 0 & \lambda_r \end{bmatrix} = A + B + C \tag{40}$$

A is the rotation part, B is the shear part and C is the stretching part

Then, we are going to analyze how these traditional vortex identification methods are contaminated.

### 4.1 Contamination of Q method

The scalar magnitude of Q method can be calculated based on velocity gradient tensor in the Principal Coordinates.

$$\nabla \vec{V} = \begin{bmatrix} \lambda_{cr} & -\frac{1}{2}R & 0 \\ \frac{1}{2}R + \varepsilon & \lambda_{cr} & 0 \\ \xi & \eta & \lambda_r \end{bmatrix} = \begin{bmatrix} \lambda_{cr} & \frac{1}{2}\varepsilon & \frac{1}{2}\xi \\ \frac{1}{2}\varepsilon & \lambda_{cr} & \frac{1}{2}\eta \\ \frac{1}{2}\xi & \frac{1}{2}\eta & \lambda_r \end{bmatrix} + \begin{bmatrix} 0 & -\frac{1}{2}R - \frac{1}{2}\varepsilon & -\frac{1}{2}\xi \\ \frac{1}{2}R + \frac{1}{2}\varepsilon & 0 & -\frac{1}{2}\eta \\ \frac{1}{2}\xi & \frac{1}{2}\eta & 0 \end{bmatrix} = A_Q + B_Q \tag{41}$$

$$Q = \frac{1}{2}\left(\|B_Q\|_F^2 - \|A_Q\|_F^2\right)$$

$$= \frac{1}{2}\left[2\left(\frac{R}{2} + \frac{\varepsilon}{2}\right)^2 + 2\left(\frac{\xi}{2}\right)^2 + 2\left(\frac{\eta}{2}\right)^2\right] - \frac{1}{2}\left[2\lambda_{cr}^2 + \lambda_r^2 + 2\left(\frac{\varepsilon}{2}\right)^2 + 2\left(\frac{\xi}{2}\right)^2 + 2\left(\frac{\eta}{2}\right)^2\right]$$

$$= \left(\frac{R}{2}\right)^2 + \frac{1}{2}R \cdot \varepsilon - \lambda_{cr}^2 - \frac{1}{2}\lambda_r^2 \tag{42}$$

It can be clearly seen that from the expression of Q above there is not only R, which is the magnitude of rotation, but also $\varepsilon$, $\lambda_{cr}$ and $\lambda_r$ which are either the shear part or stretching part in the Q-criterion. Therefore, the value of Q is certainly contaminated by shear and stretching.

## 4.2 Contamination of $\lambda_{ci}$ Criterion

The characteristic equation of velocity gradient tensor given by (18) is

$$(\lambda - \lambda_r)\left[(\lambda - \lambda_{cr})^2 + \frac{R}{2\left(\frac{R}{2}+\varepsilon\right)}\right] = 0 \tag{43}$$

Thus, the eigenvalues are

$$\lambda_1 = \lambda_r, \lambda_2 = \lambda_{cr} + i\sqrt{R/2(R/2 + \varepsilon)}, \lambda_3 = \lambda_{cr} - i\sqrt{R/2(R/2 + \varepsilon)}$$

Since rotation is orthogonal, eigenvalues are the same as the original velocity gradient tensor,

$$\lambda_2 = \lambda_{cr} + i\sqrt{R/2(R/2 + \varepsilon)} = \lambda_{cr} + i\lambda_{ci}$$

$$\lambda_3 = \lambda_{cr} - i\sqrt{R/2(R/2 + \varepsilon)} == \lambda_{cr} - i\lambda_{ci}$$

Therefore, we have

$$\frac{R}{2}\left(\frac{R}{2} + \varepsilon\right) = \lambda_{ci}^2 \tag{44}$$

Thus,

$$\lambda_{ci} = \sqrt{\frac{R}{2}\left(\frac{R}{2} + \varepsilon\right)} \tag{45}$$

The expression of $\lambda_{ci}$ has $\varepsilon$, which is in the shear part of the Principal Decomposition. As a result, $\lambda_{ci}$ is contaminated by shear.

## 4.3 Contamination of $\Delta$ method

In section 4.2, it is known that three roots of the characteristic equation is

$$\lambda_1 = \lambda_r, \lambda_2 = \lambda_{cr} + i\sqrt{R/2(R/2 + \varepsilon)}, \lambda_3 = \lambda_{cr} - i\sqrt{R/2(R/2 + \varepsilon)}$$

Plug their values into (2.),(3.) and (4.)

$$I_1 = -(\lambda_1 + \lambda_2 + \lambda_3) = -\lambda_r - 2\lambda_{cr} \tag{46}$$

$$I_2 = \lambda_1\lambda_2 + \lambda_2\lambda_3 + \lambda_3\lambda_1 = 2\lambda_r\lambda_{cr} + \lambda_{cr}^2 + \frac{R}{2}\left(\frac{R}{2} + \varepsilon\right) \tag{47}$$

$$I_3 = -\lambda_1\lambda_2\lambda_3 = -\lambda_r\left[\lambda_{cr}^2 + \frac{R}{2}\left(\frac{R}{2} + \varepsilon\right)\right] \tag{48}$$

$$\tilde{Q} = I_2 - \frac{1}{3}I_1^2 = -\frac{1}{3}(\lambda_{cr} - \lambda_r)^2 + \frac{R}{2}\left(\frac{R}{2} + \varepsilon\right) \tag{49}$$

$$\tilde{R} = I_3 + \frac{2}{27}I_1^3 - \frac{1}{3}I_1I_2 = \frac{2}{27}(\lambda_{cr} - \lambda_r)^3 + \frac{2}{3}(\lambda_{cr} - \lambda_r)\frac{R}{2}\left(\frac{R}{2} + \varepsilon\right) \tag{50}$$

Then, the expression of $\Delta$ can be written as

$$\Delta = \left(\frac{\tilde{Q}}{3}\right)^3 + \left(\frac{\tilde{R}}{2}\right)^2 = \frac{1}{243}\left[9\left(\frac{R}{2}\right)^3\left(\frac{R}{2} + \varepsilon\right)^3 - 6\left(\frac{R}{2}\right)^2\left(\frac{R}{2} + \varepsilon\right)^2(\lambda_{cr} - \lambda_r)^2 + \frac{5R}{2}\left(\frac{R}{2} + \varepsilon\right)(\lambda_{cr} - \lambda_r)^4\right] \tag{51}$$

Obviously, $\varepsilon$, $\lambda_r$ and $\lambda_{cr}$ are included in the expression of $\Delta$, which indicates that $\Delta$ is contaminated by shear and stretching.

## 5. Vortex Example: Burge Vortex

A test case of a real vortex, namely Burger vortex, is examined to justify the comparison of the effects of shearing and stretching/compression on different criteria. The velocity gradient tensor has been obtained from the Burger vortex, which is an exact steady solution of the Navier-Stokes equation and can be used to model fine scales of turbulence (Gao et al., 2019)[28]. The Burger vortex forms when an inward radial flow concentrates and spins around the symmetric axis, and the flow moves out in both directions along the z-axis (Webster, D. R., and Young, D. L., 2015)[29,30].

The velocity components of Burger vortex in the cylindrical coordinate system is given by:

$$v_r = -\xi r \tag{52}$$

$$v_\theta = \frac{\Gamma}{2\Pi r}\left(1 - e^{\frac{-r^2\xi}{2\nu}}\right) \tag{53}$$

$$v_z = 2\xi z \tag{54}$$

where $\Gamma$ represents the circulation, $\xi$ the axisymmetric strain rate, $\nu$ the kinematic viscosity, and $r$ is the distance of the chosen point from the centerline in the Burger vortex.

For post-processing, the velocity components are converted into the Cartesians coordinate system given by:

$$u = -\xi x - \frac{\Gamma}{2\Pi r^2}\left(1 - e^{\frac{-r^2\xi}{2\nu}}\right) y \tag{55}$$

$$v = -\xi y + \frac{\Gamma}{2\Pi r^2}\left(1 - e^{\frac{-r^2\xi}{2\nu}}\right) x \tag{56}$$

$$w = 2\xi z \tag{57}$$

The existence of the vortex structure is highly contingent on the selection of parameters. For the proper vortex structure visualization, we take $\xi = 0.042$, $\Gamma = 1.45$ and $\nu = 0.01$. The calculation domain is taken with $50 \times 20 \times 20$ grid points with a step size of 0.5. The streamlines of Burger vortex exhibit a spiral pattern around the vortex rotation axis line, which has the maximum vortex strength. The streamlines representing such a flow are demonstrated in Figs. 1a and 1b, which show that flows enter from radial direction and stretches outward spinning around the axis. The vortex strength is strong in the core and becomes weak when moving away from the center. We can see this phenomenon in the following figures:

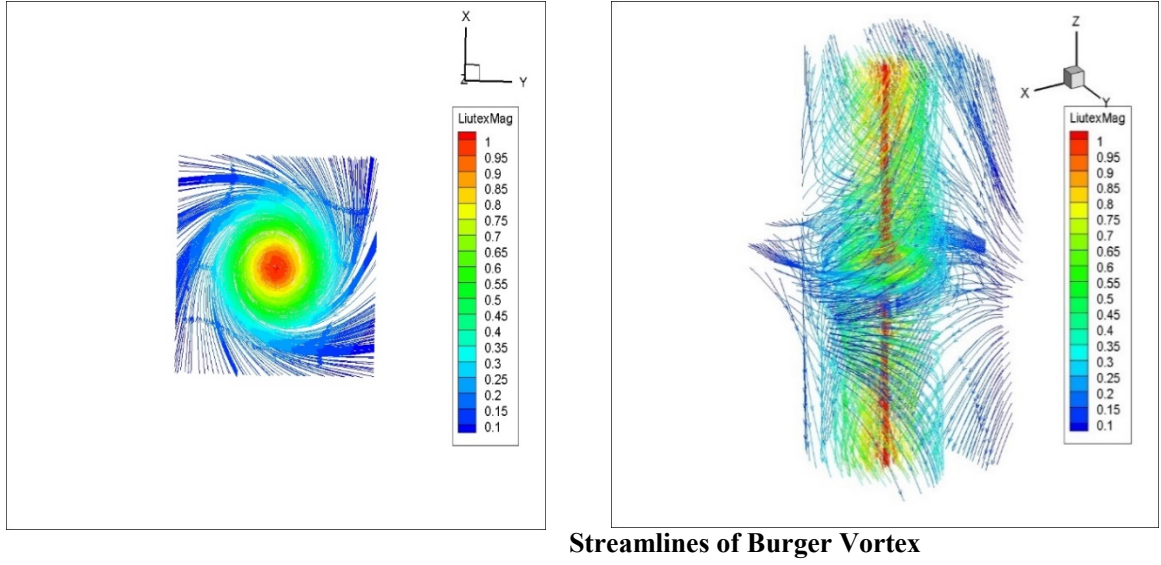

**Streamlines of Burger Vortex**

(a) Top View　　　　　　　　　　　　　　　　(b) Side View

Fig 1. (Color online) Streamlines of Burger vortex from top and side view with Liutex magnitude, which depicts the rotational strength of the fluid particles.

## 6. Numerical Contamination Analysis

The velocity gradient tensor in the Principal Coordinate is:

$$\nabla V = \begin{bmatrix} \frac{\partial u}{\partial x} & \frac{\partial u}{\partial y} & \frac{\partial u}{\partial z} \\ \frac{\partial v}{\partial x} & \frac{\partial v}{\partial y} & \frac{\partial v}{\partial z} \\ \frac{\partial w}{\partial x} & \frac{\partial w}{\partial y} & \frac{\partial w}{\partial z} \end{bmatrix} = \begin{bmatrix} \lambda_{cr} & -\frac{1}{2}R & 0 \\ \frac{1}{2}R + \varepsilon & \lambda_{cr} & 0 \\ \xi & \eta & \lambda_r \end{bmatrix} \quad (58)$$

where $u$, $v$ and $w$ are the three components of the velocity along $x$, $y$, and $z$ directions in Cartesian Coordinate.

Since our purpose is to demonstrate the shear and stretching contaminations of different criterion numerically, we add shear and stretching components separately and calculate how different criterion responds.

### 6.1 Adding Shear components

The matrix corresponding to shear is in the form of

$$\nabla V_{shear} = \begin{bmatrix} 0 & 0 & 0 \\ \varepsilon_a & 0 & 0 \\ \xi_a & \eta_a & 0 \end{bmatrix} \qquad (59)$$

The subscript "$a$" refers to "adding" of shearing components.

The new velocity gradient tensor in the Principal Coordinates after adding shear is

$$\nabla V_1 = \begin{bmatrix} \lambda_{cr} & -\frac{1}{2}R & 0 \\ \frac{1}{2}R + \varepsilon + \varepsilon_a & \lambda_{cr} & 0 \\ \xi + \xi_a & \eta + \eta_a & \lambda_r \end{bmatrix} \qquad (60)$$

Under Principal Coordinate, the local rotation axis is z-axis, so $\xi_a$ (value of $\frac{\partial w}{\partial x}$ component) and $\eta_a$ (value of $\frac{\partial w}{\partial y}$ component) are not in the rotation plane. Thus, these two issues will not influence the rotation strength.

However, $\varepsilon_a$ is in the rotation plane, and it is possible to affect rotation strength. By Liutex definition, the magnitude of rotation strength is $\min_\theta \left\{ \left|\frac{\partial u}{\partial y}\right|_\theta, \left|\frac{\partial v}{\partial x}\right|_\theta \right\}$. $\left|\frac{\partial u}{\partial y}\right|_\theta$ and $\left|\frac{\partial v}{\partial x}\right|_\theta$ are the minimum absolute values of $\frac{\partial u}{\partial y}$ and $\frac{\partial v}{\partial x}$ respectively, when we rotate the coordinate $\theta$ angle anti-clockwise along z-axis.

**Proposition 1**: The rotation strength does not change if $\varepsilon_a \geq -\varepsilon$; the rotation strength changes if $\varepsilon_a < -\varepsilon$

Proof:

If $\varepsilon_a \geq -\varepsilon$, then $\left|\frac{1}{2}R + \varepsilon + \varepsilon_a\right| \geq \left|-\frac{1}{2}R\right|$,

So, the Liutex magnitude is still $R$ which is defined as twice of the angular speed of the rigid rotation

Remark: Physically, $\varepsilon_a < 0$ means adding a shear against the rotation direction and $\varepsilon_a < -\varepsilon$ indicates that shear is strong to the extent that it can affect rotation strength.

So, keeping the rotation strength the same, the shear matrix we add should satisfy the following requirement:

$\varepsilon_a \geq -\varepsilon$.

## 6.2 Adding Stretching components

The matrix corresponding to stretching is of the form

$$\nabla V_{stretching} = \begin{bmatrix} \alpha_a & 0 & 0 \\ 0 & \beta_a & 0 \\ 0 & 0 & \gamma_a \end{bmatrix} \tag{61}$$

The subscript "$a$" means "adding" the stretching components.

The new velocity gradient tensor after adding stretching is

$$\nabla V_2 = \begin{bmatrix} \lambda_{cr} + \alpha_a & -\frac{1}{2}R & 0 \\ \frac{1}{2}R + \varepsilon & \lambda_{cr} + \beta_a & 0 \\ \xi & \eta & \lambda_r + \gamma_a \end{bmatrix} \tag{62}$$

**Proposition 2**: The rotation strength does not change if $\alpha_a = \beta_a$.

Proof: It is easy to know $\nabla V_2$ is the velocity gradient tensor under the Principal Coordinate. Based on theorem 1, its rotation strength is still $R$.

Therefore, to keep the rotation strength unchanged, the added $\nabla V_{stretchin}$ should satisfy $\alpha_a = \beta_a$.

## 6.3 Stretching Contamination Analysis

The following procedure was implemented to depict how the vortex identification schemes react over the change in stretching effects. Firstly, we select a point in Burgers vortex, and its velocity gradient tensor is:

$$\nabla \vec{V}_B = \begin{bmatrix} -0.0419999994 & -0.0711665452 & 0.0000000000 \\ 0.0711665452 & -0.0419999994 & 0.0000000000 \\ 0.0000000000 & 0.0000000000 & 0.0839999988 \end{bmatrix} \tag{63}$$

Then for our convenience, the stretching matrix $\nabla V_{stretchin}$ is added and the values of $Q, \Delta, \lambda_2, \lambda_{ci}$, and Liutex are recorded where $\nabla V_{stretching}$ is given by (note that it satisfies the requirement of adding stretching):

$$\nabla V_{stretchin} = \begin{bmatrix} 0.02 & 0 & 0 \\ 0 & 0.02 & 0 \\ 0 & 0 & -0.04 \end{bmatrix} \tag{64}$$

The sum of diagonal elements of $\nabla V_{stretching}$ is equal to zero, so it satisfies the continuity equation. The stretching effect given by $\nabla V_{stretchin}$ is increased repeatedly, and the values of every scheme is recorded. The results are presented in the following graph, where the x-axis represents relative stretching rate, which is the ratio of stretching component and vorticity magnitude, and the y-axis gives the corresponding values of different vortex identification methods.

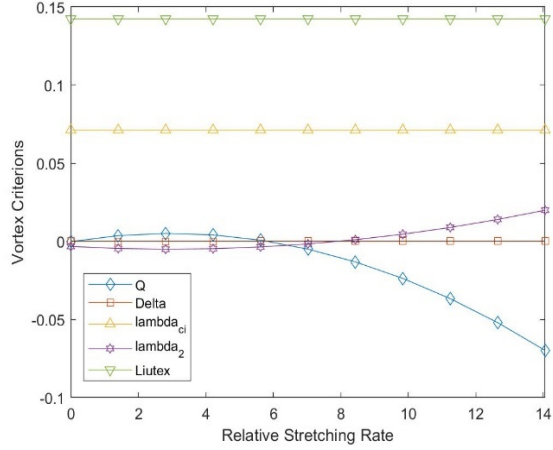

Fig. 2. Line graphs depicting the effect of stretching on different vortex identification methods.

The change in stretching components result in a change in $Q$, $\Delta$, $\lambda_2$, but $\lambda_{ci}$ and L are not changed, which means $\lambda_{ci}$ and L are not contaminated by stretching effect, as shown in Fig. 2. It can also be observed that increasing the stretching effect would significantly decrease the positive value of $Q$ and ultimately becomes negative indicating non-vortical structure. However other criterions still show that there is a vortex structure. It can be concluded that $Q$ may conflict with other criteria. Theoretically, this makes sense as in equation (5), $\Delta$ could still be positive even if $Q$ is negative provided that the square of the third variant is large enough. The computational results shown in Fig. 2 are coincided with theoretical analysis in Eqs. (42) and (45).

## 6.4 Shear Contamination Analysis

A similar procedure to the stretching effect is implemented for the graphic representation of the shearing effect on the vortex identification schemes. Again, for our convenience, the shearing matrix $\nabla V_{shear}$ is defined as:

$$\nabla V_{shear} = \begin{bmatrix} 0 & 0 & 0 \\ 0.02 & 0 & 0 \\ 0.02 & 0.02 & 0 \end{bmatrix} \tag{65}$$

$\nabla V_{shear}$ is added to $\nabla \vec{V}_B$ for few times, and then the corresponding values of the different criterions are recorded. The results are presented in the following graph, where the x-axis represents relative shear rate, and the y-axis gives the corresponding values of different vortex identification methods.

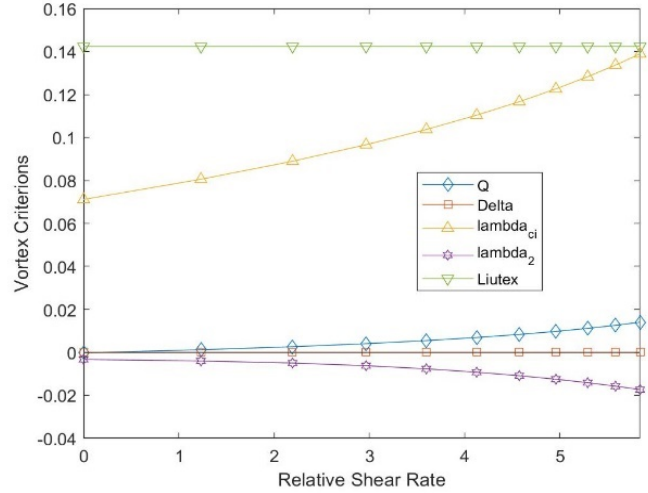

Fig. 3. Line graphs depicting the effect of shear on different vortex identification methods.

The relative shear rate is the ratio of the shearing component over vorticity magnitude. More precisely, it is the ratio of shear over vorticity at any point in our numerical domain. The reason we are dividing shear by vorticity is to refrain shear from getting too large. Fig. 3 indicates that $Q$, $\Delta$, $\lambda_2$, and $\lambda_{ci}$ are all greatly affected by the change in the shearing component, while L has remained unaffected by shear. It can be concluded from Fig. 2 and Fig. 3 that $Q$, $\Delta$, $\lambda_2$, and $\lambda_{ci}$ are all affected by either shear, stretching, or both at different levels whereas the Liutex method is affected by neither.

Since the values of $\Delta$ are very small, $\Delta$ looks to be consistently zero; however, if we use zoom in, it can be observed that the $\Delta$ values are increasing. The $\Delta$ - plots are given in Fig. 4 below:

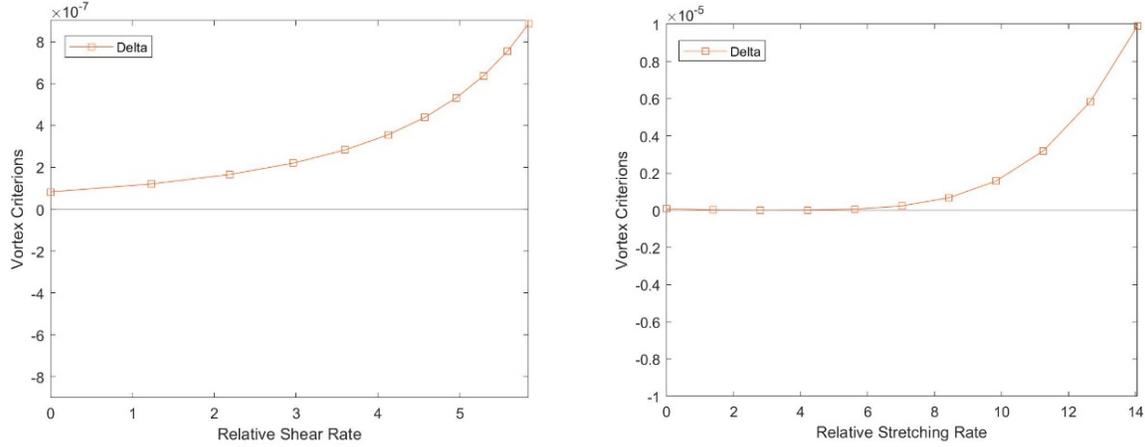

Fig. 4. Stretching and shearing effect on $\Delta$ method of vortex identification.

## 7. Conclusion

As the theoretical relation between Liutex and several second generation of vortex identification methods is given and a numerical example of Burger vortex is implemented, the following conclusions can be summarized.

1) Mathematical relation between Liutex and the second generation of vortex identification methods, such as $Q$, $\Delta$, and $\lambda_{ci}$ methods, is derived by mathematical formula, which clearly shows the second generation methods are seriously contaminated by shearing or stretching or by both while Liutex is the exact and mathematical definition of fluid rotation or vortex.

2) Velocity gradient tensor in the Principal Coordinates can be uniquely decomposed into the rigid rotational part, stretching/compression part, and shearing part, which is also known as Liutex decomposition.

3) $Q$ conflicts with $\Delta$, $\lambda_2$, and $\lambda_{ci}$ methods. At some point, Q may mistreat the vortex point as a non-rotational point which can be seen in the graphical representations too (see Figure 2).

4) According to Table 1, $Q$, $\Delta$, and $\lambda_2$ are stretch sensitive, while $Q$, $\Delta$, $\lambda_2$, and $\lambda_{ci}$ are affected by the shearing factor. The numerical results from the Burger vortex coincide with the theoretical analysis. However, Liutex remains the same and is not affected. Therefore, we can conclude $Q$, $\lambda_2$, and $\lambda_{ci}$ methods are contaminated by shear or stretching or both, and may not be appropriate to represent vortex strength while Liutex is the only one which correctly represents fluid rotation or vortex.

Table 1. Contamination by stretching and shearing on different criteria.

| Methods | $Q$ | $\Delta$ | $\lambda_2$ | $\lambda_{ci}$ | L |
|---|---|---|---|---|---|
| Contamination by stretching | Yes | Yes | Yes | No | No |
| Contamination by shearing | Yes | Yes | Yes | Yes | No |

## 8. Data availability statement

The data that supports the findings of this study are available from the corresponding author upon reasonable request.